\documentclass{llncs}
\usepackage{makeidx}  
\usepackage{graphicx}  
\usepackage{multirow}
\usepackage{cite}
\usepackage{subfigure}
\usepackage{amsmath}
\usepackage[utf8]{inputenc}
\usepackage[numbers, square, comma, sort&compress]{natbib}
\begin{document}
\frontmatter          
\pagestyle{headings}  
\addtocmark{Hamiltonian Mechanics} 
\title{OMG Emotion Challenge - ExCouple Team}
\titlerunning{Hamiltonian Mechanics}  
%

\author{Ingryd Pereira$^{1}$ \and Diego Santos$^{2}$}

%
%
%
\institute{$^{1}$Polytechnic School of Pernambuco, University of Pernambuco, Recife, Brazil\\
$^{2}$ Federal University of Pernambuco, Recife, Brazil}

\maketitle              

%
The proposed model is just for the audio module. 
The expression of emotion through the voice is one of the essential forms of human communication \cite{weninger2013acoustics}.
Although the voice is a reliable affection source, recognize the affection through the voice is a more complicated task \cite{liu2015emotional, schuller2011recognising}.

One of the challenges when processing audio, is the representation of the audio characteristics. For a long time, handmade transformations have stood out in this area, such as the MFCC. But traditional feature extraction techniques loss too much information from the audio unlike deep learning models which possible the use of the lowest level of raw speech, like spectral characteristics, for speech recognition and automatically learns to make this transformation. Deep learning models also allow the use of convolutional and clustering operations to represent and deal with some typical speech variability (e.g., differences in vocal tract length at high-speakers, different speech styles, etc.)\cite{deng2014deep}.


But deep learning models require a high number of labeled training data to perform well, and there is a scarcity of emotional data available, which makes the task of emotion recognition challenging.

The semi-supervised learning can overcome the lack information problem of labeled data. For the OMG Challenge, we use a GAN, which has unsupervised learning,  to learn and generate the audio representation and this representation will be used as input for the model that will predict the values of arousal and valence.
The benefit of using this approach is that part of the model that will represent the audio can be trained with any database, with a much larger amount of data, since it does not require a label for your training. Doing this also creates a general model of audio representation, which allow the use of the model in different tasks and different databases without retraining.

To develop the application used for this challenge, we use a BEGAN that uses an autoencoder as a discriminator. The encoder part of this autoencoder learns how to perform the audio representation. For the BEGAN training, we use the IEMOCAP database, which is one of the largest emotional databases available. The training occurred in 100 epochs, with batch size 16, and with a $\gamma$ value of 0.7.

We only use the audio module from the database, but all files are available in mp4 video format. So as preprocessing the application extracts and saves the audio from all videos in the database as WAV format. The next step is to change the audio frequency to 16kH. Then each audio track was decomposed into 1-second chunks without overlapping. After that, the raw audio was converted to a spectrogram via Short Time Fourier Transform, with an FFT of size 1024 and a length of 512.

\begin{figure}[!ht]
\centering
\includegraphics[scale=0.35]{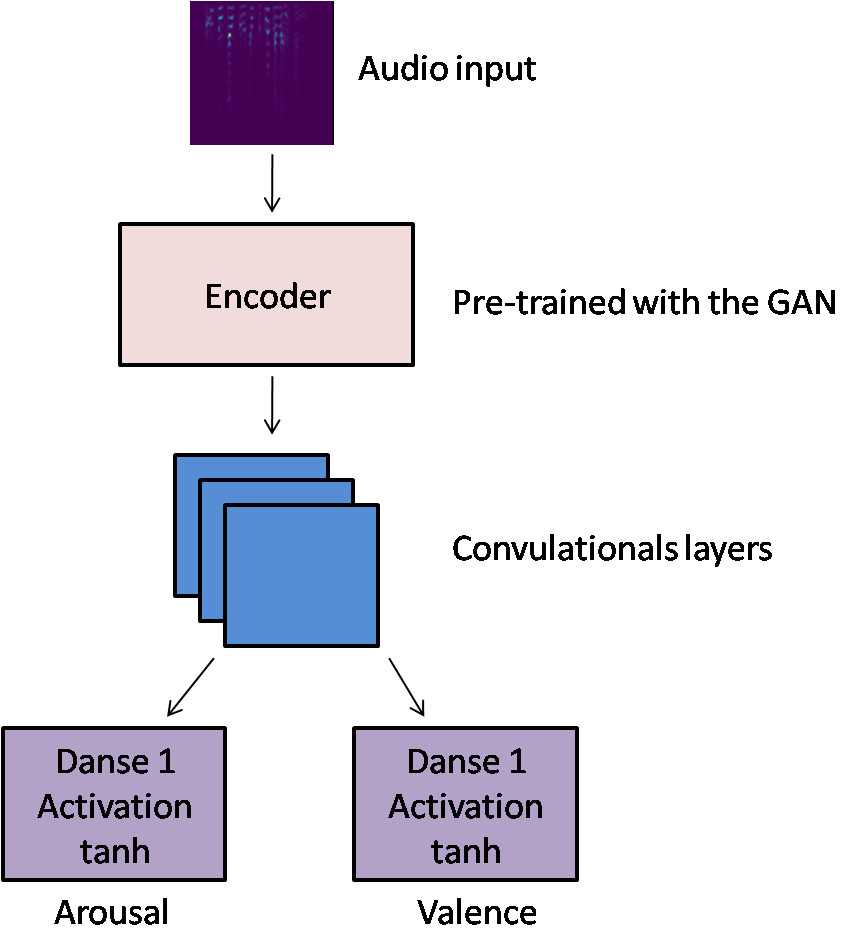}
\caption{Abstraction of the classifier and prediction models}
\label{abstraction}
\end{figure}

Figure \ref{abstraction} presents the developed model abstraction. The model uses the preprocessed audio as input to the representation module pre-trained by the BEGAN. The encoder output is the input for a set of convolutional layers followed by dense layers with activation \textit{tahn}, which predicts the arousal and valence values (values between -1 and 1).

In preprocessing, the application divides the audios into 1-second pieces, performing the prediction for each of these pieces. But at the end of the prediction process, it is necessary to gather the results from each part and check out the value of arousal and valence for whole audio. To do this, we use the median value of the predicted values of each 1-second part from given audio. The median of the set of predicted arousal values will be the representation of arousal of that given audio, and the application uses the same process for valence value.

\begin{figure}[!ht]
\centering
\includegraphics[scale=0.35]{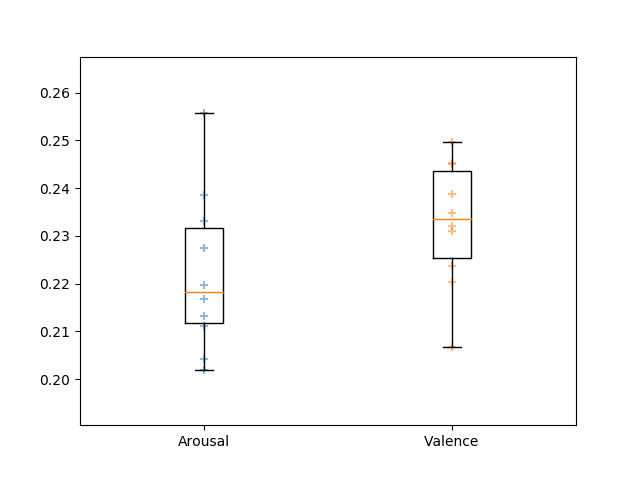}
\caption{Box plot with the CCC of the Arousal and Valence values predicts}
\label{boxplot}
\end{figure}

Figure \ref{boxplot} shows the box-plot with the predicted arousal and valence values in 10 model runs. The base line present in Barros et al. work, \cite{barros2018omg} has a CCC better than 0.15 and valence 0.21, and as can be seen in Figure \ref{boxplot}, our model obtain better results.

%
%



\begin{thebibliography}{1}
\bibitem{liu2015emotional}Liu, Mengmeng and Chen, Hui and Li, Yang and Zhang, Fengjun \textsl{"Emotional tone-based audio continuous emotion recognition"}. International Conference on Multimedia Modeling. Springer. 2015

\bibitem{schuller2011recognising} Schuller, Bj{\"o}rn and Batliner, Anton and Steidl, Stefan and Seppi, Dino \textsl{"Recognising realistic emotions and affect in speech: State of the art and lessons learnt from the first challenge"}. Speech Communication. Elsevier. 2011

\bibitem{deng2014deep} Deng, Li and Yu, Dong and others \textsl{"Deep learning: methods and applications"}. Foundations and Trends{\textregistered} in Signal Processing. Now Publishers, Inc.. 2014

\bibitem{weninger2013acoustics} Weninger, Felix and Eyben, Florian and Schuller, Bj{\"o}rn W and Mortillaro, Marcello and Scherer, Klaus R \textsl{"On the acoustics of emotion in audio: what speech, music, and sound have in common"}. Frontiers in psychology. Frontiers Media SA. 2013



\bibitem{barros2018omg} Barros, Pablo and Churamani, Nikhil and Lakomkin, Egor and Siqueira, Henrique and Sutherland, Alexander and Wermter, Stefan \textsl{"The OMG-Emotion Behavior Dataset"}. arXiv preprint arXiv:1803.05434. 2018

\bibliography{mybib,paper}
\end{thebibliography}
\end{document}